\documentclass[11pt]{article}
\usepackage{fullpage,graphicx,psfrag,amsmath,amsfonts,verbatim}
\usepackage{xcolor}
\usepackage{amsthm}
\usepackage[small,bf]{caption}
\usepackage{authblk}

\usepackage{amsmath}
\usepackage{amssymb}
\usepackage{mathtools}
\usepackage{amsthm}
\usepackage{amsfonts}
\usepackage{algorithm}
\usepackage{algorithmic}

\usepackage{tabularx}
\usepackage{makecell}
\usepackage{multirow}

\usepackage{microtype}
\usepackage{graphicx}
\usepackage{booktabs} 
\usepackage{hyperref}
\usepackage{subcaption}
\usepackage{enumitem}

\allowdisplaybreaks

\title{Generative Multi-Agent Collaboration in Embodied AI: A Systematic Review}

\author[1]{Di Wu}
\author[3]{Xian Wei}
\author[1]{Guang Chen}
\author[4]{Hao Shen}
\author[3]{Xiangfeng Wang}
\author[1]{\\Wenhao Li\thanks{\texttt{whli@tongji.edu.cn}}}
\author[2,1]{Bo Jin\thanks{\texttt{bjin@tongji.edu.cn}}}

\affil[1]{School of Computer Science and Technology, Tongji University}
\affil[2]{Shanghai Research Institute for Intelligent Autonomous Systems, Tongji University}
\affil[3]{School of Computer Science and Technology, East China Normal University}
\affil[4]{Technical University of Munich}
\begin{document}
\maketitle
\begin{abstract}
Embodied multi-agent systems (EMAS) have attracted growing attention for their potential to address complex, real-world challenges in areas such as logistics and robotics. 
Recent advances in foundation models pave the way for generative agents capable of richer communication and adaptive problem-solving. 
This survey provides a systematic examination of how EMAS can benefit from these generative capabilities. 
We propose a taxonomy that categorizes EMAS by system architectures and embodiment modalities, emphasizing how collaboration spans both physical and virtual contexts. 
Central building blocks, perception, planning, communication, and feedback, are then analyzed to illustrate how generative techniques bolster system robustness and flexibility. 
Through concrete examples, we demonstrate the transformative effects of integrating foundation models into embodied, multi-agent frameworks. 
Finally, we discuss challenges and future directions, underlining the significant promise of EMAS to reshape the landscape of AI-driven collaboration.
\end{abstract}

\section{Introduction}


Embodied multi-agent systems (EMAS) have garnered growing interest due to their significant potential in domains such as smart transportation, logistics, and manufacturing~\cite{yan2013survey,ismail2018survey}. 
By integrating physical embodiments—ranging from autonomous vehicles to robotic manipulators—with multi-agent systems (MAS)~\cite{dorri2018multi}, EMAS offers a decentralized, collaborative approach that can handle complex tasks with remarkable efficiency. 
Despite these advantages, designing and implementing effective EMAS remains a non-trivial endeavor, often requiring specialized knowledge of cybernetics, extensive training data, and careful reinforcement-learning paradigms~\cite{landauer2008computational,orr2023multi}.


In traditional MAS, agents collaborate by dividing responsibilities, sharing state information, and collectively adapting to dynamic environments~\cite{dorri2018multi}. 
While these principles have led to impressive success in certain niches, conventional approaches face critical limitations in generalizing to new tasks~\cite{mahajan2022generalization}, scaling to large agent populations~\cite{cui2022survey}, and coping with unexpected environmental changes~\cite{weinberg2004best}. 
They often rely on narrowly trained models, which can be brittle or constrained to specific domains~\cite{yuan2023robust}. 
These shortfalls underscore the urgency for more flexible and robust solutions that can thrive in open-ended and rapidly changing embodied scenarios.


Recent breakthroughs in foundation models (FMs, e.g., large language models, FMs, or vision-language models, VLMs)~\cite{zhou2024comprehensive} have opened new avenues for advancing MAS toward more adaptive and generative behaviors. 
By equipping agents with natural language capabilities, contextual reasoning, and the ability to generate novel solutions, FM-based MAS transcend some of the limitations inherent in purely signal-driven or reinforcement-based frameworks~\cite{guo2024large,chen2024survey,lu2024merge} . 
These ``generative agents'' can communicate in semantically rich ways, collaborate with human-level fluency, and rapidly adapt strategies in response to unforeseen challenges. 
As such, FM-powered agents could transform how multi-agent collaboration unfolds—both in physical spaces populated by embodied devices and in virtual spaces where agents share abstract knowledge and tasks.

Against this backdrop, the field of EMAS stands poised to benefit from these recent advances in FMs.
By combining physical embodiments with generative multimodal intelligence, future systems may embrace a broader design space that incorporates complex perception, high-level linguistic and visual reasoning, and adaptive decision-making. 
However, existing literature surveys on embodied AI and multi-agent systems often treat these fields in isolation, leaving critical gaps at their intersection~\cite{ismail2018survey,duan2022survey,guo2024large,ma2024survey,hunt2024survey}.
A systematic view of how FM-based generative agents can best be integrated into EMAS is still emerging.



\begin{figure}[htb!]
\centering
    \includegraphics[width=\linewidth]{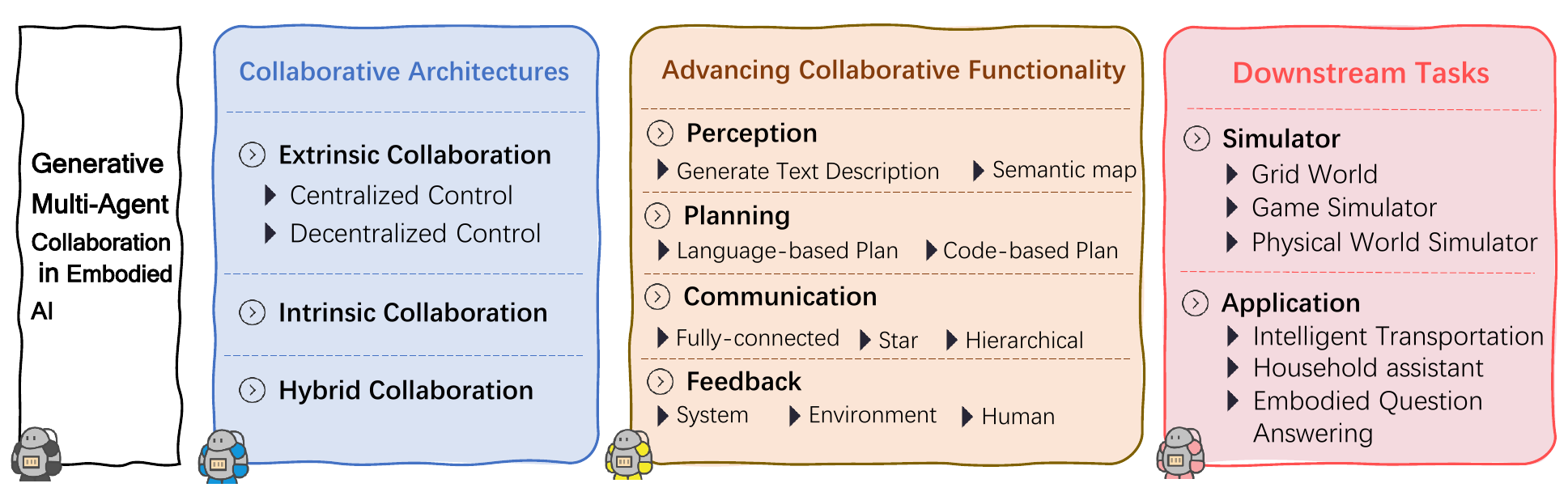}
    \caption{A unified multi-agent framework for generative embodied AI.}
    \label{fig:overall_large_pic}
\end{figure}


This survey aims to provide a comprehensive and structured examination of the current state of generative multi-agent collaboration in Embodied AI, as shown in Figure~\ref{fig:overall_large_pic}. 
First, we introduce a taxonomy that classifies existing EMAS solutions based on the number of models and types of embodiments in Section~\ref{sec:arch}, highlighting how collaboration can arise both among physical agents and in purely virtual semantic environments.
Next, we explore the major building blocks of multi-agent collaboration—system perception, planning, communication, and feedback—and examine how each of these components can be designed to leverage FM-based generative capabilities in Section~\ref{sec:modular}. 
Moving beyond theoretical perspectives, we delve into practical applications in Section~\ref{sec:app}, illustrating how generative multi-agent collaboration enhances functionality across diverse embodied scenarios.


To the best of our knowledge, this is the first survey to systematically address the convergence of MAS, Embodied AI, and foundation models. 
We close by summarizing open research challenges in Section~\ref{sec:future}, delineating crucial future directions, and discussing the potential impact of EMAS on broader AI and robotics landscapes.
Our goal is to inform and inspire researchers, practitioners, and stakeholders by presenting a holistic overview of this rapidly evolving field.

\begin{figure}[htb!]
\centering
    \includegraphics[width=0.7\linewidth]{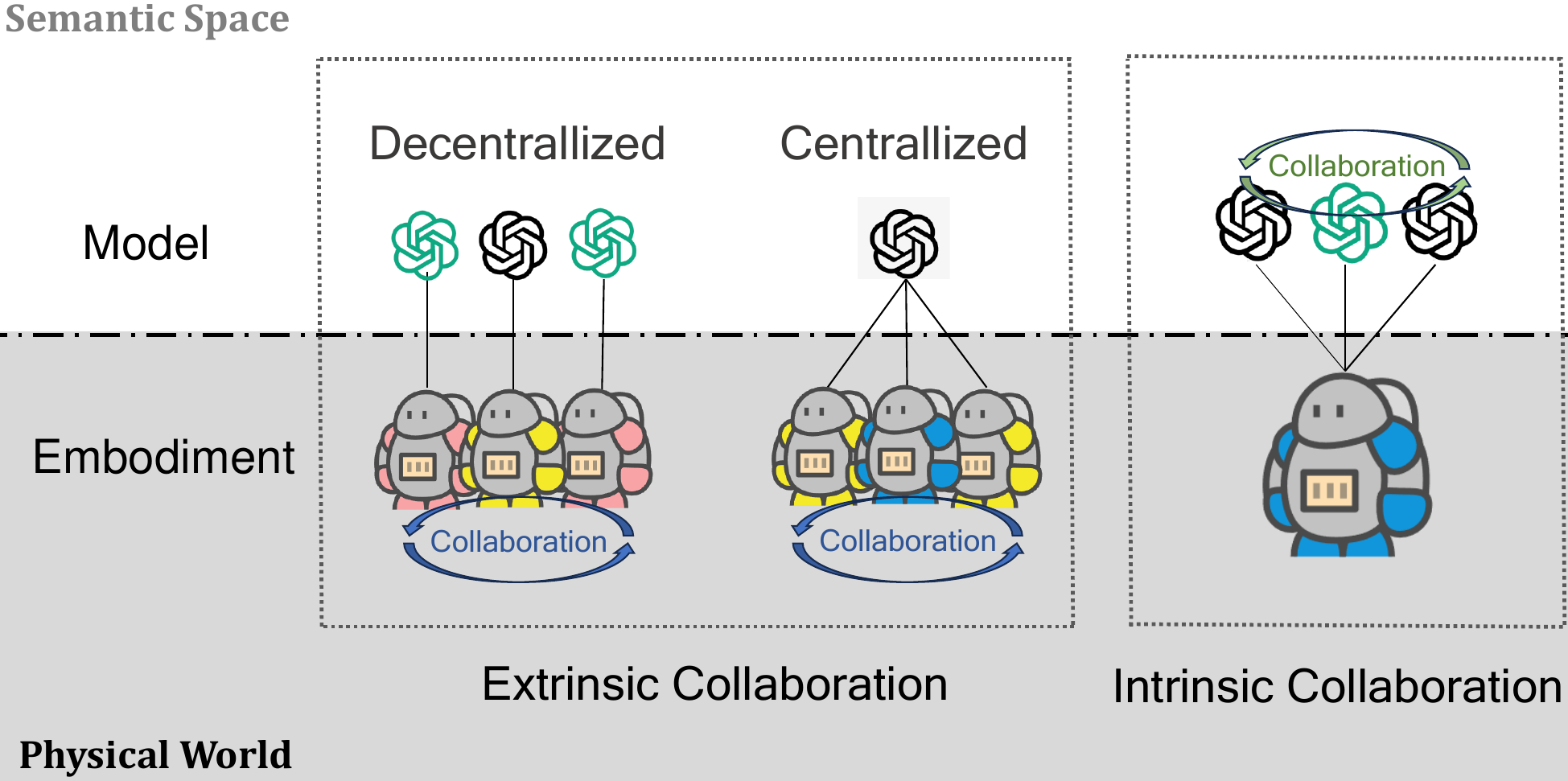}
    \caption{The embodied multi-agent collaborative architecture.}
    \label{fig:architecture}
\end{figure}

\section{Collaborative Architectures}\label{sec:arch}

Building upon the key challenges and opportunities outlined in the previous section, this section presents an overview of collaborative architectures in EMAS, as shown in Figure~\ref{fig:architecture}.
In particular, we address how generative multi-agent systems leverage either \emph{extrinsic collaboration} across multiple embodied entities or \emph{intrinsic collaboration} among multiple FMs within a single embodied entity. 
We also cover \emph{hybrid} approaches that combine these strategies to meet diverse system requirements. 
The goal is to provide a structured understanding of how multi-agent collaborations can be orchestrated to maximize adaptability, scalability, and task alignment, especially when integrated with FMs.


\subsection{Extrinsic Collaboration}

In scenarios where collaboration unfolds among multiple embodied entities, known as \emph{extrinsic collaboration}, agents physically or virtually interact to accomplish shared objectives. 
Drawing from the longstanding multi-robot and conventional MAS literature, extrinsic collaboration can be organized using either a centralized or a decentralized strategy. 
These approaches offer differing trade-offs related to scalability, communication overhead, and global versus local control.

\paragraph{Centralized Architecture}
In a centralized policy framework, a single unified model controls multiple robots or agents, offering centralized task allocation and decision-making. 
The centralized model assigns tasks based on agent capabilities and system objectives, ensuring coordination across agents by providing a global perspective. 
Studies have explored language-based~\cite{liu_coherent_2024,obata_lip-llm_2024,chen_emos_2024} and code-based~\cite{kannan_smart-llm_2024,zhang_lamma-p_2024} task allocation.

The centralized model also plays a key role in decision-making by synthesizing information from all agents to make final decisions, ensuring coherence. 
For example, \cite{yu_co-navgpt_2023} uses a centralized model for navigation target determination, \cite{tan_multi-agent_2020} uses it for interactive question answering with a 3D-CNN-LSTM QA model, and \cite{garg_foundation_2024} employs it for deadlock resolution in multi-robot systems by guiding a leader robot’s actions.

The centralized control strategy ensures coordination by using a single model for task allocation and decision-making. 
Its advantages include optimal task distribution and consistent decisions. 
However, it can be limited by system complexity, high computational demands, and scalability issues in large or dynamic environments.

\begin{table*}[htb!]
\renewcommand{\arraystretch}{1.25}
\setlength{\tabcolsep}{3pt}
\tiny
\begin{tabularx}{\linewidth}{c|c|p{2cm}|X|X|X|X|X|X|X}
\hline
\multicolumn{2}{c|}{\textbf{Collaboration Architecture}} & 
\textbf{Work} & 
\multicolumn{5}{c|}{\textbf{Component}} & 
\multicolumn{2}{c}{\textbf{Application}} \\ \cline{4-10}

& & & 
{\color[HTML]{036400} \textbf{Perception}} & 
{\color[HTML]{F56B00} \textbf{Planning}} & 
{\color[HTML]{3166FF} \textbf{Comm.}} & 
{\color[HTML]{F400FE} \textbf{Feedback}} & 
{\color[HTML]{4D4D4D} \textbf{Agent}}    &
{\color[HTML]{003366} \textbf{Env.}} & 
{\color[HTML]{4D4D4D} \textbf{Task}} \\ \hline

\multirow{18}{*}{\makecell{Extrinsic\\Collaboration}} & 
\multirow{7}{*}{Centralized} & 
\cite{yu_co-navgpt_2023} & Semantic map & Language & Star & -- & homo-agents  & Simulator & Navigation \\ \cline{3-10}
& & \cite{liu_coherent_2024} & Description & Language & Star & Environment & hetero-agents & Simulator, real-world & Household tasks \\ \cline{3-10}
& & \cite{chen_emos_2024} & Semantic map & Language & FC,Star & System & hetero-agents & Simulator & Household tasks \\ \cline{3-10}
& & \cite{kannan_smart-llm_2024} & Description  & Code & Star & -- & homo-agents & Simulator, real-world & Household tasks \\ \cline{3-10}
& & \cite{zhang_lamma-p_2024} & -- & Code (PDDL) & Star & System & hetero-robots & Simulator & Household tasks \\ \cline{3-10}
& & \cite{tan_multi-agent_2020} & Semantic map & -- & Star & -- & homo-agents & Simulator & EQA \\ \cline{3-10}
& & \cite{wang_safe_2024} & Description & Language & Star & Human & homo-agents & Simulator & Household tasks \\ \cline{2-10}

& \multirow{11}{*}{Decentralized} & 
\cite{mandi_roco_2023} & Visual detect & Language & FC & Environment & homo-agents & Simulator, real-world & Household tasks \\ \cline{3-10}
& & \cite{zhang_building_2024} & Semantic map & Language & FC & -- & homo-agents & Simulator & Household tasks \\ \cline{3-10}
& & \cite{zhang_combo_2024} & World model & Language & -- & System & homo-agents & Simulator & Cooperative, Competitive game \\ \cline{3-10}
& & \cite{agashe_llm-coordination_2024} & Descrpition & Language & -- & System & homo-agents & Game & Overcooked, hanabi game \\ \cline{3-10}
& & \cite{guo_embodied_2024} & -- & Language & Hierarchical & Environment & homo-agents, multi-role llm & Simulator & Household tasks \\ \cline{3-10}
& & \cite{chen_s-agents_2024} & Description & Language & Hierarchical & -- & homo-agents & Game & Minecraft creation \\ \cline{3-10}
& & \cite{chen_autotamp_2024} & Description & Language & -- & System & homo-agents, multi-role llm & Grid & Multi grid tasks \\ \cline{3-10}
& & \cite{chen_agentverse_2024} & -- & Language & Hierarchical & Environment & homo-agents & Game & Minecraft creation \\ \cline{3-10}
& & \cite{zhang_towards_2024} & -- & Language & FC & System & homo-agents & Simulator,game & Household tasks, overcooked game \\ \cline{3-10}
& & \cite{tan_knowledge-based_2023} & Semantic map & -- & -- & -- & homo-robots & Simulator & EQA \\ \cline{3-10}
& & \cite{wu_hierarchical_2024} & Description & Language & Hierarchical & Environment & homo-agents & Game & Minecraft navigation \\ \hline

\multirow{6}{*}{\makecell{Intrinsic\\Collaboration}} & & 
\cite{shi_opex_2024} & Semantic Map & Language & -- & -- & multi-role llm & Simulator & Household tasks \\ \cline{3-10}
& & \cite{qin_mp5_2024} & Description & Language & -- & Environment & multi-role llm & Game & Minecraft creation, navigation \\ \cline{3-10}
& & \cite{nayak_long-horizon_2024} & Description & Language & -- & System, Environment & multi-role llm & Simulator & Household tasks \\ \cline{3-10}
& & \cite{zhang_controlling_2024} & Description & -- & Hierarchical & System & multi-role llm & Grid & Grid Transportation \\ \cline{3-10}
& & \cite{fastandslow} & Description & Language & -- & System,Human & multi-role llm & Game & Overcooked game \\ \cline{3-10}
& & \cite{chen_towards_2023} & Visual detect & Language & -- & System & multi-role llm & -- & EQA \\ \hline

\end{tabularx}
\caption{Taxonomy of representative works. ``--'' denotes that a particular element is not specifically mentioned in this work.}
\end{table*}

\paragraph{Decentralized Architecture}

In a decentralized strategy, each model independently controls its corresponding embodied entity, providing greater flexibility and scalability. 
Early studies used reinforcement learning for decentralized control, but the rise of FMs has enabled agents to handle diverse tasks autonomously~\cite{chen_s-agents_2024}, forming more advanced decentralized systems.

FMs enhance decentralized systems by leveraging the reasoning capabilities to improve individual decision-making based on local partial observations. 
For example, \cite{zhang_combo_2024} utilizes a world model to assist multi-agent planning, where each individual predicts the behavior of other agents through the world model and infers its own plan. 
Similarly, \cite{agashe_llm-coordination_2024} introduces an auxiliary theory-of-mind reasoning FM to interpret the actions and needs of partner agents, thereby supporting individual decision-making.

Furthermore, with the reasoning and communication capabilities of FMs, FM-based agents exhibit emergent sociality. 
\cite{chen_multi-agent_2023} reveals that when not explicitly instructed on which strategy to adopt, FM-driven agents primarily follow the average strategy, representing an egalitarian organizational structure among agents. 
Other research~\cite{guo_embodied_2024,chen_s-agents_2024} highlight the potential benefits of more structured roles within the team. 
This suggests that, similar to human social structures, FM agents can exhibit emergent behaviors that optimize collaboration by adapting to organizational frameworks, enhancing their collective ability to tackle complex tasks.

\subsection{Intrinsic Collaboration}


While extrinsic collaboration deals with multiple robots and embodied entities, \emph{intrinsic collaboration} occurs within the internal structure of a single system that may contain multiple FMs. 
This concept resonates with the recent push for collaborative workflows among various FM modules, each specializing in different roles, to jointly handle increasingly complex tasks. 
Such internal orchestration expands traditional notions of multi-agent coordination by focusing on consolidated decision-making within a single embodiment.

Each FM in this workflow assumes a specific function or \emph{role} to collaboratively complete a task. 
Research has applied this paradigm to embodied learning systems, such as~\cite{qin_mp5_2024}, which uses modules like planner, partoller, and performer for task-solving in a Minecraft sandbox, and \cite{shi_opex_2024}, which decomposes tasks into observer, planner, and executor roles. 
LLaMAR~\cite{nayak_long-horizon_2024} also employs a plan-act-correct-verify framework for self-correction without oracles or simulators.

Intrinsic collaboration can improve system functionality by enhancing planning accuracy, safety, and adaptability. 
For example, \cite{fastandslow} uses FM-based fast-mind and slow-mind for collaborative plan generation and evaluation, while LLaMAC~\cite{zhang_controlling_2024} employs multiple critics and an assessor to provide feedback and improve robustness.

\subsection{Hybrid Collaboration Architectures}

In many real-world applications, drawing a strict boundary between extrinsic and intrinsic collaboration is neither practical nor advantageous. 
Instead, hybrid collaborative architectures combine these strategies to exploit the strengths of centralized, decentralized, and internal FM workflows. 
As embodied tasks grow in complexity, the flexibility to mix different levels of collaboration, both among robots and within an agent’s internal structure, becomes increasingly valuable.


Intrinsic collaboration enhances model capabilities through modular FMs and can be applied in centralized and decentralized systems. 
For example, CoELA~\cite{zhang_building_2024} uses five modules--perception, memory, communication, planning, and execution--while~\cite{yu_mhrc_2024} builds agents with observation, memory, and planning modules for decentralized robot collaboration. 
Centralized models can also use modular FMs, such as~\cite{wu_hierarchical_2024}, which employs a task- and action-FM for task assignment.


Centralized and decentralized strategies can be combined, with different stages of a task utilizing different approaches. 
Inspired by the centralized training with decentralized execution (CTDE) framework in multi-agent reinforcement learning (MARL), \cite{chen_emos_2024} and \cite{zhao_hierarchical_2024} propose centralized planning with decentralized execution, where global planning guides task execution, maximizing the synergy between global oversight and local autonomy.

By showcasing these varying architectures, we illustrate how practitioners can effectively orchestrate multi-agent collaboration in EMAS across different levels of granularity and control. 
The next section builds on this architectural perspective by examining how key system components--perception, planning, communication, and feedback--can be designed to leverage FM-based generative capabilities for more robust and adaptive multi-agent collaboration.

\section{Advancing Collaborative Functionality}\label{sec:modular}

Building upon the architectural insights from Section~\ref{sec:arch}, which examined how multi-agent collaboration can be orchestrated at the structural level, we now pivot to the functional building blocks that drive effective teamwork among embodied agents. 
Specifically, we highlight how \emph{perception}, \emph{planning}, \emph{communication}, and \emph{feedback} mechanisms can be designed to harness the generative capabilities of FMs. 
By focusing on these key modules, we illustrate how EMAS solutions can more robustly interpret the physical environment, formulate and adapt plans, exchange information, and iteratively learn from both their own behaviors and the environment itself. 
This approach complements the collaboration architectures introduced previously, offering a finer-grained perspective on enabling dynamic and context-aware cooperation among embodied agents. 

\subsection{Perception}

Although a generative model may derive semantic knowledge from text and vision, embodied agents must actively sense and interpret the physical world. 
This entails handling 3D structures, dynamic conditions, and real-time interactions~\cite{liu2024aligning}. 
Consequently, the perception module is paramount, as it conveys detailed environmental features to subsequent models, ensuring that generative capabilities are grounded in tangible contexts~\cite{pan2024recent}. 

\paragraph{Physical Perception for FM}


The simplest means of providing physical context to an FM is to supply a verbal description of the environment. 
Although such prompts may be crafted manually, many approaches augment linguistic descriptions with automated tools. 
For instance, some studies~\cite{mandi_roco_2023,chen_towards_2023} use visual models to detect and describe objects, while others~\cite{brohan2023can,huang2023voxposer} employ affordance learning to enrich the FM’s understanding of how objects can be operated upon in a physical setting.
Beyond passively receiving information, recent work enables agents to decide \emph{when} and \emph{what type} of information to observe, facilitating active perception. 
For example, \cite{qin_mp5_2024} allows the FM to query a fine-tuned model about environmental details; the responses inform a progressively constructed scene description.

\paragraph{Collaborative Perception}


In multi-agent systems, collaborative perception aims to merge complementary sensory inputs from different agents, enhancing overall performance~\cite{yang_spatio-temporal_2023}. 
Within autonomous driving or drone fleets, this often arises through sensor-level data sharing or output-level fusion~\cite{singh2024multi}. 
In FM-based systems, collaborative agents may collectively build a global memory of the environment by aggregating each agent’s local maps or visual data. 
For instance, \cite{yu_co-navgpt_2023} fuses the semantic maps derived from RGBD inputs of multiple agents, and \cite{tan_multi-agent_2020} employs 3D reconstruction of each agent’s observations to form a holistic 3D status and semantic memory of the shared environment.

\subsection{Planning}


Planning constitutes a core module of multi-agent embodied systems, enabling agents to strategize based on states, goals, and individual capabilities. 
Effective planning is crucial for task assignment, coordination, and seamlessly integrating the capabilities of generative FMs.

\paragraph{Planning format}
Planning methods often employ either \emph{language-based} or \emph{code-based} formats. 
Language-based planning uses natural language to guide task flows, achieving intuitiveness and ease-of-adaptation, especially with the advent of advanced FMs~\cite{mandi_roco_2023,yu_co-navgpt_2023}. 
By contrast, code-based methods utilize structured programming or domain-specific notations (e.g., PDDL) for higher precision. 
\cite{kannan_smart-llm_2024} uses Python code to frame overall task flow, and \cite{zhang_lamma-p_2024} converts tasks into PDDL problems for allocation to multiple robots.

\paragraph{Planning process}
Beyond individual decision-making, multi-agent collaboration demands consensus-building, conflict resolution, and resource sharing. 
In \emph{centralized} systems, a single model frequently allocates sub-tasks. 
For example, \cite{liu_coherent_2024} generates action lists based on each agent’s capability, \cite{obata_lip-llm_2024} integrates FMs and linear programming to solve task partitions, while \cite{chen_emos_2024} exploits ``robot resumes'' for FM-based discussions around task assignment. 
In \emph{decentralized} systems, agents communicate directly to optimize their collective plans, supported by robust information exchanges that will be explored in the following subsection.

\subsection{Communication}

Communication is central to MAS, enabling agents to share situational updates, coordinate tasks, and reach consensus. 
Unlike traditional approaches that require painstaking communication protocol design, generative agents can exploit the zero-shot language generation abilities of FMs, reducing the complexity of building efficient communication interfaces.

Following~\cite{sheng2022learning}, we categorize multi-generative-agents communication patterns in embodied AI to three main structures:
\begin{itemize}[leftmargin=*]
    \item \textbf{Star}: A virtual central agent controls the flow of messages, broadcasting plans or directives to other agents.Much work with centralized architecture has explored this approach~\cite{kannan_smart-llm_2024,yu_co-navgpt_2023}
    \item \textbf{Fully Connected (FC)}: Every agent communicates freely with every other agent, leveraging FM-driven messages. For instance, \cite{mandi_roco_2023} uses inter-FM dialogues between two robotic arms to coordinate manipulation tasks. In CoELA~\cite{zhang_building_2024}, each agent references current state information via memory retrieval, generating communication content through an FM.
    \item \textbf{Hierarchical}: A leadership structure emerges to boost scalability and reduce communication overhead. \cite{chen_s-agents_2024,liu_agentlite_2024,guo_embodied_2024} show how leadership roles channel or filter communications, improving efficiency and outcomes.
\end{itemize}

\subsection{Feedback}

Embodied tasks are complex and uncertain, making feedback mechanisms essential for agent improvement. 
Feedback enables agents to adjust and optimize behavior, allowing continuous learning based on the current state, environmental changes, or external guidance. 

\paragraph{System Feedback}
System feedback refers to information generated internally before an action is taken. 
This involves agents or a centralized model revisiting their initial plans to identify flaws or potential improvements. 
Several works~\cite{liu_capo_2024,chen_emos_2024,zhang_controlling_2024} implement a multi-agent discussion phase post-plan generation, refining action lists through peer feedback. 
\cite{chen_autotamp_2024} and \cite{zhang_lamma-p_2024} employ FM checkers to validate code-based plans, ensuring syntactic correctness. 
Meanwhile, \cite{zhang_towards_2024} devises advantage functions to evaluate and iteratively refine plans, and \cite{fastandslow} applies an FM to predict plan consequences, followed by another FM that rates plan quality, thus driving iterative enhancements.

\paragraph{Environmental Feedback}

Environmental feedback surfaces after executing actions in the physical (or simulated) world. 
Many studies log real-world outcomes to guide future decisions. 
For example, \cite{liu_coherent_2024} and \cite{yu_mhrc_2024} store action results in memory for future planning references, whereas \cite{qin_mp5_2024} and \cite{nayak_long-horizon_2024} assess the root cause of failures and adapt their action plans accordingly. 
Additionally, multi-agent organizational structures can be reconfigured mid-task in response to environmental signals. 
\cite{chen_agentverse_2024} dynamically updates the roles, and \cite{guo_embodied_2024} employs a critic FM to evaluate agent performance, even reorganizing leadership.

\paragraph{Human Feedback}


External human guidance can offer nuanced interventions and strategic directions unattainable through purely automated systems. 
For instance, \cite{park2023clara} identifies \emph{ambiguous} or \emph{infeasible} task instructions warranting human assistance, while \cite{wang_safe_2024} and \cite{ren2023robots} integrate conformal prediction to measure task uncertainty and trigger human help requests. 
Beyond soliciting assistance, \cite{cui2023no} and \cite{shi2024yell} permit human operators to refine on-the-fly robot actions through spoken instructions, improving task success rates.

In sum, perception, planning, communication, and feedback emerge as foundational pillars for translating high-level collaborative architectures into practical, generative multi-agent solutions. 
Whether agents collaborate extrinsically through distributed configurations or intrinsically via multiple FMs within a single embodied, robust supporting modules ensure adaptability and resilience in real-world settings. 

The next section delves into concrete application domains, illustrating how these functional modules synergize to tackle diverse embodied tasks. 
By bridging architectural principles (Section~\ref{sec:arch}) with modular functionalities and grounding them in real-world scenarios, we aim to offer a comprehensive view of how generative multi-agent collaboration can be effectively realized in EMAS.

\section{Downstream Tasks: From Simulation to Real-World Deployment}\label{sec:app}


Building on the architectures and functional modules, this section examines how generative multi-agent collaboration moves from controlled simulation environments to real-world applications. 
Although many advances are validated through virtual platforms, these simulation insights lay the groundwork for tackling the complexities of intelligent transportation, household robotics, and embodied question answering.


\subsection{Simulation Platforms}\label{sec:simulators}

Earlier sections introduced how multi-agent collaboration can be structured and functionally enabled. 
Simulation environments now enter as a crucial layer for testing these designs, allowing researchers to systematically refine agent interactions without incurring real-world operational costs or risks. 

\paragraph{Grid-World Paradigms}
Grid worlds feature cell-based structures that focus on decision-making and path planning while abstracting away physical details. 
By adopting an FM-based translator-and-checker framework, \cite{chen_autotamp_2024} improves multi-agent performance on grid tasks, while \cite{zhang_controlling_2024} introduces feedback mechanisms to enhance grid transportation. 
\cite{chen_scalable_2024} further evaluates various FM-driven multi-robot architectures in a grid setup, underscoring how these simplified worlds facilitate quick validation of collaborative designs.

\begin{table}[ht]
\renewcommand{\arraystretch}{1.25} 
\setlength{\tabcolsep}{2pt} 
\small
\begin{tabularx}{\linewidth}{X|X|p{5.4cm}}
\cline{1-3} 
\textbf{Simulator Level} & \textbf{\begin{tabular}[c]{@{}l@{}}Environment \\ and Benchmark\end{tabular}} & \textbf{Related paper} \\ \cline{1-3} 

\textbf{Grid} & Grid World & \cite{zhang_controlling_2024}, \cite{chen_autotamp_2024}, \cite{chen_scalable_2024} \\ \cline{1-3} 
\multirow{2}{*}{\textbf{Game}} & Minecraft & \cite{chen_s-agents_2024}, \cite{park_mrsteve_2024}, \cite{qin_mp5_2024}, \cite{zhao_hierarchical_2024}, \cite{zhao_we_2024} \\ \cline{2-3} 

& Overcooked-AI & \cite{agashe_llm-coordination_2024}, \cite{ying_goma_2024}, \cite{zhang_towards_2024} \\ \cline{1-3} 
 
\multirow{8}{*}{\textbf{Advanced 3D}} & ThreeDWorld & \cite{zhang_combo_2024}, \cite{liu_capo_2024}, \cite{zhang_building_2024} \\ \cline{2-3} 

& AI-THOR & \cite{kannan_smart-llm_2024}, \cite{wang_safe_2024}, \cite{nayak_long-horizon_2024}, \cite{zhang_lamma-p_2024}, \cite{liu_heterogeneous_2023}  \\ \cline{2-3} 
 
& ALFRED & \cite{shi_opex_2024} \\ \cline{2-3} 
& Habitat & \cite{chen_emos_2024}, \cite{yu_co-navgpt_2023} \\ \cline{2-3} 
& VitualHome & \cite{ying_goma_2024}, \cite{guo_embodied_2024} \\ \cline{2-3} 
& RocoBench & \cite{mandi_roco_2023}, \cite{zhang_towards_2024} \\ \cline{2-3} 
& Bestman & \cite{yu_mhrc_2024} \\ \cline{2-3} 
& Behavior-1k & \cite{liu_coherent_2024} \\ \cline{1-3} 
\end{tabularx}
\caption{Simulators and applications.}
\label{tab:my-table-3}
\end{table}

\paragraph{Game-Based Collaboration Scenarios}
Game-based platforms like Overcooked provide clear rules, time constraints, and forced coordination among agents~\cite{ying_goma_2024,agashe_llm-coordination_2024,zhang_towards_2024}. 
FM-coordination extends to other structured games such as Hanabi and Collab Games, showcasing that generative approaches are adaptable to diverse team-based challenges.
For more open-ended tasks, Minecraft~\cite{wang_voyager_2023,park_mrsteve_2024} pushes the envelope with larger environments and indefinite goals. 
Recent work~\cite{park_mrsteve_2024,zhao_hierarchical_2024,qin_mp5_2024} focuses on collaborative exploration, while others~\cite{chen_s-agents_2024,chen_agentverse_2024,zhao_we_2024} tackle resource collection or structure building.

\paragraph{Advanced 3D Environments and Robotic Simulations}

Realistic simulators aim to mirror real-life complexity more closely. 
AI2-THOR~\cite{Kolve2017AI2THORAn} offers meticulously rendered indoor scenes and is used for multi-agent household tasks~\cite{kannan_smart-llm_2024,wang_safe_2024,liu_embodied_2022,shi_opex_2024}. 
Similarly, VirtualHome-Social~\cite{guo_embodied_2024}, BEHAVIOR-1K~\cite{liu_coherent_2024}, and Habitat-based benchmarks~\cite{chen_emos_2024} enable agents to develop collaborative strategies in object manipulation and navigation. 
Such platforms help bridge the gap between algorithmic development and physical deployment.



\subsection{Emerging Applications}\label{sec:emerge-app}

Armed with validated architectures and robust functional modules, researchers have begun to face the ultimate frontier: translating simulator learnings into viable physical deployments. 
From intelligent transportation to household robotics, the following subsections spotlight how generative multi-agent collaboration is being adapted to meet real-world demands, illustrating both the maturity and remaining challenges of these systems.

\paragraph{Intelligent Transportation and Delivery}
Multi-agent collaboration in intelligent transportation covers UAV/UGV coordination for cargo delivery and environmental monitoring. 
Early approaches mainly leveraged MARL, but FM-driven solutions are now emerging. 
\cite{gupte_rebel_2024} explores FM-based initial task allocation for surveillance missions, and \cite{wu_hierarchical_2024} applies generative models to assign tracking targets, suggesting that language-guided strategies can adapt swiftly to dynamic scenarios.

\paragraph{Household Assistance Robotics}
Many 3D simulation benchmarks, including AI2-THOR and Habitat, were originally crafted to emulate domestic environments. 
Domestic tasks such as ``clearing the table'' or following instructions like ``Turn on the desk and floor lamp and watch TV'' demand robust perception, planning, and communication. 
Studies~\cite{kannan_smart-llm_2024,wang_safe_2024,liu_heterogeneous_2023,mandi_roco_2023,zhang_towards_2024} demonstrate how multiple agents can share roles, interpret commands, and divide complex tasks. 
Generative models further streamline coordination, enabling adaptive task assignment and richer human-robot interactions.

\paragraph{Beyond Exploration: Embodied Question Answering}
Embodied Question Answering (EQA) involves active exploration and reasoning in 3D spaces. 
Unlike tasks that emphasize physical interactions, EQA focuses on gathering and interpreting information, often requiring an advanced understanding of spatial layouts, object relationships, or event histories.
Multi-agent versions often leverage team-based sensing for global memory and consensus~\cite{tan_knowledge-based_2023,tan_multi-agent_2020,patel_multi-llm_2024}. 
\cite{chen_towards_2023} positions agents with specialized functions to contribute key information, showcasing how FM-driven collaboration can integrate observations into coherent answers.

In highlighting these simulation benchmarks and real-world applications, this section underscores a key trajectory in EMAS: leveraging structured testbeds for proof-of-concept, then transitioning solutions to high-stakes domains. 
Having established where and how generative multi-agent collaboration can be deployed, the subsequent sections will address remaining challenges and outline prospective frontiers for EMAS research.

\section{Open Challenges and Future Trends}\label{sec:future}

As the field of multi-agent collaboration in embodied AI systems is continuing to develop, there remain several open challenges and promising future directions. 
Despite the progress made, numerous real-world obstacles persist, limiting the application of embedied systems. 
This section identifies key challenges and outlines potential areas of exploration and innovation to address these issues.

\paragraph{Benchmarking and Evaluation}
One major challenge is lacking standardized evaluation criteria. 
While significant strides have been made in benchmarking individual agents and single-agent systems, there is a notable gap for the evaluation of embodied multi-agent collaboration. 
Existing benchmarks focus on task-specific metrics, and fail to account for the complexity of interactions, coordination, and emergent behaviors, that arise in multi-agent settings. 
There is an urgent need for unified evaluation standards for the holistic performance, including factors such as scalability, adaptability, robustness, and collective intelligence. 
The development of benchmarks is crucial to ensure consistent across different domains, and enabling meaningful comparisons between various multi-agent frameworks.

\paragraph{Data Collection and Heterogeneity}
Another challenge in multi-agent collaboration is the data scarcity and heterogeneity for embodied systems.
Collecting large-scale, high-quality datas of different systems with diverse physical characteristics and capabilities is an arduous task. 
The variation in hardware, sensor, and environmental interactions leads to inconsistence, making it difficult to generalize across systems and tasks. 
The real-world data available could be limited, hindering training and evaluating effectively. 
Additionally, most works in multi-agent collaboration are conducted in simulated environments, due to practical constraints, and only a few studies employ real-world data. 
Hence, there is a pressing need for standardized data collection, as well as innovative methods to transfer between simulation and real-world applications, to bridge the gap between theory and reality.

\paragraph{Foundation Models for Embodied AI}
The development of foundation models, particularly for embodied agents, is poised to be a transformative breakthrough in the field of multi-agent collaboration. 
Currently, generative agents primarily rely on FMs to perform complex tasks, and naturally the next step is to build specifical foundational models designed for embodied systems. 
These models serve as a core framework for multi-agent collaboration, integrating perception, decision-making, and action. 
Recent works, such as RT-1~\cite{brohan2022rt} and RDT~\cite{liu2024rdt}, 
made significant strides in robot foundation models for adaptable and scalable systems. 
The evolution of foundation models will lay the groundwork for more seamless multi-agent collaboration, enabling agents with comprehensive capabilities and teamworks in dynamic environments. 
However, challenges remain to extend single-agent FMs to multi-agent, requiring novel architectures and methodologies.

\paragraph{Scalability of Agents}
Currently, numbers of agents involved in collaborative multi-agent  systems remains small. 
Scaling up the number of agents will lead to the increased complexity and difficulty of computation, communication, coordination, task allocations, and resource management. 
Moreover, maintaining stability and robustness in large-scale multi-agent systems requires sophisticated orchestration and coordination techniques. 
Researches on scalable architecture, efficient communication protocol, and collaborative tactics 
will be essential to unlocking the full potential of large-scale embedied systems. 
The development to optimize agent workflows and patterns will be crucial for scaling up these systems in a resource-awareness manner.

\paragraph{Human-Centric Collaboration}
The integration of robots into human-centered environments remains a critical topic. 
In many applications, multi-agent systems need to collaborate with not only each other but also  human. 
Ensuring that robots can work seamlessly alongside humans in dynamic and unstructured environments requires the development of human-robot interaction (HRI) protocols that consider human cognitive capabilities, preferences, and limitations. 
Human-robot collaboration introduces additional challenges, such as safety, adaptability, and trustworthy. 
Researches on human-robot teamwork, shared autonomy, and intuitive interfaces will be vital for fostering productive and safe collaboration between humans and robots, particularly for healthcare, industrial automation, and service robots.

\paragraph{Theoretical Foundations and Interpretability}
Current approaches of embodied multi-agent collaborations, particularly those involving FMs, often lack a solid theoretical foundation. 
While substantial progress has been made in developing practical systems, the understanding is very limited about the underlying principles and collective intelligence that emerges to govern agent interactions. 
A deeper theoretical exploration of the dynamic cooperation, including the roles of communication, coordination, and consensus, is essential for advancing the field. 
Furthermore, the reliability and interpretability of embedied multi-agent systems and models is critical, especially for safety-critical environments, such as automatic drive and smart railway.


\section{Related Work}

Although numerous surveys have examined embodied AI or MAS individually, few efforts have tackled the critical overlap between these fields, leaving significant knowledge gaps unaddressed.
Early studies on embodied AI~\cite{hu2023toward,firoozi2023foundation,ma2024survey} focus on single-agent perception-action loops. 
They discuss autonomy and sensorimotor learning in depth, yet devote limited attention to collaborative paradigms. 
Similarly, \cite{duan2022survey} and \cite{liu2024aligning} explore how agents interact with environments but still assume solitary agents with limited capacity for distributed teamwork.

In contrast, recent surveys on FM-driven multi-agent systems~\cite{guo2024large,chen2024survey,lu2024merge} showcase promising results in semantic communication and emergent coordination, especially in virtual environments. 
However, these contributions remain detached from physical embodiment, where hardware constraints, sensor noise, and kinematic coordination pose significant challenges. 
Meanwhile, classic robotics surveys~\cite{ismail2018survey,yan2013survey} laid the groundwork for cooperative swarm behaviors but lack generative capabilities that facilitate role adaptation or zero-shot planning.

Several specialized reviews provide partial bridges. 
For instance, \cite{hunt2024survey} advances language-based human-robot interaction, yet overlooks non-linguistic coordination crucial in industrial or warehouse settings. 
Likewise, \cite{sun2024llm} integrates FMs with MARL but treats embodiment mostly as an implementation detail. 
Consequently, none of these views examine physical grounding, collaborative intelligence, and generative models under a unified lens.

Our survey addresses this gap by synthesizing insights from three converging axes: 
(i) embodied AI’s physical imperatives, 
(ii) multi-agent systems’ collaborative intelligence, and 
(iii) generative models’ adaptive reasoning. 
We propose a novel taxonomy that reconciles embodiment multiplicity (physical agents) with model multiplicity (virtual agents). 
Through case studies in cross-modal perception and emergent communication, we show how FMs overcome conventional multi-agent limitations in real-world embodied contexts. 
Finally, we identify underscored challenges, such as Sim2Real transfer for generative collectives, bridging the divide between robotics and FM-based coordination. 
Our work thus establishes conceptual foundations for a new generation of embodied systems, where physical constraints and generative collaboration progress in tandem.




\section{Conclusion}\label{sec:conclusion}

This survey investigates a popular and potential research area, i.e. multi-agent collaboration in embodied systems, which focuses on how generative foundation models can be integrated into embodied multi-agent systems.
We emphasize how FM-based generative agents facilitate dynamic collaboration and emergent intelligence, and systematically explore multi-agent collaboration architectures from both intrinsic and extrinsic perspectives, focusing on key technologies such as perception, planning, communication and feedback mechanisms.
Various applications range from grid world exploration to household assistance in embodied scenarios are studied to demonstrate the potential of FM-based EMAS to address complex problems, and discuss the associated challenges and opportunities in this rapidly evolving field.
We hope this survey can serve as a valuable lamp for researchers,  practitioners, and stakeholders, that offers a comprehensive understanding of the current landscape and inspires more advanced and scalable solutions of dynamic seamless collaboration for embodied multi-agent AI. 

\clearpage
\newpage
\bibliography{ijcai24}


\end{document}